\documentclass{iopart}

\begin{document} \title{LISA Sources in Globular Clusters}

\author{M J Benacquista\dag, S Portegies Zwart\footnote[7]{Hubble Fellow}\ddag, F A Rasio\ddag}

\address{\dag\ Dept.\ of Sciences, Montana State University-Billings, Billings, MT 59101} \address{\ddag\ Dept.\ of Physics and Center for Space Research, Massachusetts Institute of Technology, Cambridge, MA 02139}

\begin{abstract} Globular clusters house a population of compact binaries that will be interesting gravitational wave sources for LISA. We provide estimates for the numbers of sources of several categories and discuss the sensitivity of LISA to detecting these sources. The estimated total number of detectable sources ranges from $\sim 10$ to $\sim10^3$ with gravitational wave frequencies above $1\,$mHz. These sources are typically undetectable by any other means and thus offer an opportunity for doing true gravitational-wave astronomy. The detection of these sources would provide information about both binary star evolution and the dynamics of globular clusters. \end{abstract}

%\submitto{\CQG} 
%\maketitle

\section{Introduction}

Binary stars will be the most common continuous sources of gravitational radiation (GR) for LISA. The detection of these sources is enhanced for systems with short orbital periods and large component masses, favoring double compact stars, i.e., binaries containing white dwarf (WD), neutron star (NS), or black hole (BH) components. Globular clusters will be excellent hosts as they are known to contain large populations of compact objects in binaries formed through dynamical interactions (e.g., Bailyn 1995; Hut \etal\ 1992). While some of these systems have been detected directly through their electromagnetic radiation (e.g., as radio pulsars, X-ray or UV sources), most are thought to be invisible in conventional astronomy, but they could be very bright in gravitational radiation. For example, a large, dense globular cluster such as 47~Tuc may contain as many as $\sim10^4$ NS, of which $\sim10^3$ may be active radio millisecond pulsars, but only $\sim20$ are bright enough to have been detected in pulsar surveys (Camilo \etal\ 2000). Most of the detected pulsars are in binaries with low-mass WD companions.

If LISA can detect and identify some of these sources within specific globular clusters, then gravitational wave astronomy will be able to contribute significantly to our understanding of binary star evolution and stellar dynamics in dense cluster systems. Many Galactic globular clusters are close enough that GR emission from double degenerate binaries may be detected by LISA with a signal-to-noise (S/N) ratio  $>2$. At this level, LISA will have an angular resolution of a few square degrees. Individual clusters will be smaller than this, but the typical separation between clusters is much larger, so that LISA will be able to separate sources in different clusters. One of the primary sources of noise in the frequency band of cluster binaries is the Galactic disk population of close white-dwarf binaries, which is expected to produce a confusion-limited signal above the noise level of LISA (Hils and Bender 2000, Schneider \etal 2000). It may be possible to avoid this noise by looking specifically for signals originating out of the plane of the Galaxy, such as those arising from globular clusters. In this paper, we discuss the possible source populations and we present some simple methods for crudely estimating their numbers.

Since the angular resolution of LISA depends upon measuring the Doppler shift in frequency as a phase variation during the motion of the detector about the sun, the wavelength of the signal must be less than the size of LISA's orbit. This places an upper limit on the binary orbital period, $P_{\rm orb} = 2/f_{\rm gw} = 2\lambda/c \leq 4 AU/c \simeq 2000\,$s. In addition, we require that the S/N ratio be $>2$ for the binary to be detectable with the required positional accuracy.  By coincidence, this also imposes a maximum period $P_{\rm max}\simeq 2000\,$s  above which the confusion-limited noise from Galactic binaries rises significantly. Finally, we require that the systems be detached so that their period evolution is solely determined by GR, providing a ``clean'' signal. This should also improve the extraction of the signal from the confusion-limited background. These requirements limit the viable sources to double compact objects. We will not discuss other standard LISA sources such as contact binaries containing two main-sequence stars or a main-sequence star and a WD, as their signals will not be strong enough or clean enough. Integrating the in-spiral of a binary  driven by GR alone from $P_{\rm max}$ down to contact, we can write the total length of time spent in the LISA band as \begin{equation} T_{\rm LISA} \simeq 4\times10^6\,{\rm yr}\, \left(\frac{M_{\rm chirp}}{M_\odot}\right)^{-5/3} \left(\frac{P_{\rm max}}{2000\,{\rm s}}\right)^{8/3}, \end{equation} where $M_{\rm chirp} = (m_1m_2)^{3/5}/(m_1+m_2)^{1/5}$ is the chirp mass, given in terms of the binary component masses $m_1$ and $m_2$.

\section{Theoretical Estimates}

Numerical simulations of globular cluster dynamics can be used to try to estimate the numbers and properties of different classes of binary systems from first principles. Here we illustrate this approach from three recent studies using different numerical methods.

\subsection{$N$-body simulations}

The dynamical simulations of Portegies Zwart and McMillan (2000, hereafter PZM) were developed to estimate BH-BH merger rates for LIGO and VIRGO.  In their $N$-body simulations of globular cluster evolution, a component of $10\,M_\odot$ primordial BH quickly undergoes mass segregation, forming an effectively isolated subsystem at the cluster core in $\sim 10^9\,$yr. Through dynamical interactions, hard BH binaries are then formed and ejected from the cluster and the entire BH subsystem evaporates typically within $\sim 5\times10^9\,$yr.  The basic conclusion of this work is that globular clusters today should no longer contain any of their primordial BH. This explains naturally why no BH X-ray binary has ever been detected in a globular cluster, even though clusters contain many bright NS X-ray binaries (see, e.g., Deutsch \etal\ 2000). A large number of ejected BH binaries will coalesce within a Hubble time, and these may provide a significant event rate for ground-based interferometers.

However, {\it the last\/} BH binary remaining in the cluster core may not be ejected, since it can now only interact with the (much less massive) normal stars. The dynamical simulations of PZM suggest that the initial separations of typical BH binaries should be $\sim 30-300\,R_\odot$. The eccentricities are typically large, and roughly half of these binaries will coalesce through GR within the remaining $t_{\rm coal}\sim5\times10^9\,$yr of the cluster evolution to the present (it is easy to show that, for coalescing binaries, further hardening through dynamical interactions with normal cluster stars is negligible, and that ejection from the cluster is extremely unlikely). Unfortunately, with $T_{\rm LISA}\simeq 10^5\,$yr for two $10\,M_\odot$ BH (see eq.~1), even if we assume optimistically that $\sim100$ globular clusters in our Galaxy retained a coalescing BH binary, we still conclude that the number of BH-BH systems in the LISA band today is $\sim 100\, T_{\rm LISA}/t_{\rm coal} \ll 1$. Thus BH-BH binaries in clusters are not promising sources for LISA.

\subsection{Encounter Rates}

Davies (1995) calculated the production rates of various binaries through dynamical interactions using a combination of scattering experiments and static models of globular clusters. His results can be used to estimate the numbers of NS-NS, NS-WD, and WD-WD binaries with $P_{\rm orb} \leq 2000\,$s (Benacquista 1999).

The original dynamical simulation calculates three-body interaction cross-sections for a variety of interactions and outcomes and then determines encounter rates for a number of model globular clusters. Production rates of binaries were inferred from the cross-sections applied to the encounters. As a side result, the merger rates through GR induced in-spiral of binaries were produced for 15 Gyr runs. The runs were also extended until all binaries had merged. This generally took another 15 Gyr. With these two results, a merger rate can be determined through \begin{equation} \eta = \frac{(n_{\infty} - n_{15})}{15\,{\rm Gyr}} \end{equation} where $n_{\infty}$ is the number of mergers after all binaries have merged and $n_{15}$ is the number which have merged after the first 15 Gyr. The evolution of the orbital period can be determined from $\dot{P} = - k_o P^{-5/3}$ with
\begin{equation} k_o = \frac{96}{5} (2 \pi)^{8/3} \frac{G^{5/3}}{c^5} M_{\rm chirp}^{5/3}. \end{equation}
Combining these results in a number distribution given by
\begin{equation} dn = \frac{\eta}{k_o} P^{5/3} dP.
\end{equation}
Using the numbers found in Davies (1995) and scaling up to the entire globular cluster system, with a total number of stars $N_{\rm gc} \simeq 10^{7.5}$, we find 16 NS-NS binaries, 125 NS-WD binaries, and 176 WD-WD binaries in the LISA band.

\subsection{Monte Carlo Simulations}

Dynamical Monte Carlo simulations can be used to study self-consistently the evolution of binary populations within evolving globular cluster models (Rasio 2000). Rasio \etal\ 2000 used this approach to study the formation and evolution of NS-WD binaries, motivated by the recent detection of a large number of millisecond radio pulsars with WD companions in 47 Tuc (Camilo \etal\ 1999). The simulations include an evolving core of the background cluster, stellar evolution, mass segregation, and dynamical 3-body interactions. In the process of generating the appropriate populations of binary millisecond pulsars, the simulations also produce populations of NS-WD binaries in the period range of interest for LISA. There are two such binaries in the simulation of 47~Tuc (Rasio {\it et al} 2000). From this admittedly small sample it is possible to place an order of magnitude estimate on the number of such binaries in the globular clusters system. Assuming that the 2 NS-WD binaries (presently in the LISA band) for the $\sim10^6\,M_\odot$ in 47 Tuc is consistent throughout the globular cluster system, we would predict a total of $\sim60$ such binaries for the $\sim10^{7.5}\,M_\odot$ in all clusters. Note that this number is within a factor of two of the number estimated above from encounter rates.

\section{Semi-Empirical Estimates}

Alternatively, we can estimate the numbers of the various possible LISA sources by trying to relate them to certain types of binaries already observed in electromagnetic radiation. Here the best candidates are NS-WD systems, also observable as X-ray binaries, and NS-NS systems, also observable as binary radio pulsars.

\subsection{NS-WD Binaries}

The small number of observed ultra-compact X-ray binaries (UXBs) in globular clusters (Deutsch \etal\ 2000) can be used to estimate empirically the number of NS-WD binaries in the LISA band. UXBs are bright enough that we essentially see all of them in the Galactic globular cluster system. Assuming that they originate from the GR-driven in-spiral of NS-WD binaries all the way to contact (Rasio \etal\ 2000) and that the process is ongoing, we can estimate the number of progenitor systems in the LISA band by looking at the time spent by a system in both phases. Let $N_{\rm X}$ be the number of UXBs, and $T_{\rm X}$ their typical lifetime. Also, let $N_{\rm LISA}$ be the number of progenitors in the LISA band, with the corresponding $T_{\rm LISA}$ given by eq.~1. Then, if the process is stationary, we must have \begin{equation} \frac{N_{\rm X}}{T_{\rm X}} \sim \frac{N_{\rm LISA}}{T_{\rm LISA}}. \end{equation}

Of the 12 known bright X-ray binaries in clusters, 3 are known to have very short orbital periods ($<1\,$hr) and 3 are known to have long orbital periods (Deutsch \etal 1999). Thus, $3 \leq N_{\rm X} \leq 9$. The lifetime $T_{\rm X}$ is rather uncertain, depending on a number of poorly understood processes. A standard treatment of mass transfer driven by GR only would give $T_{\rm X}\sim10^9\,$yr, but other effects such as irradiation or tidal heating of the companion may well shorten this to $T_{\rm X}\sim 10^7\,$yr, which would then also provide consistency with the birthrate of binary millisecond pulsars (Applegate and Shaham 1994; Tavani 1991; Rasio \etal 2000). For a typical system containing a $1.4\,M_\odot$ NS and a $0.5\,M_\odot$ WD companion, eq.~1 gives $T_{\rm LISA}\simeq 7\times 10^6\,$yr. Adopting the most optimistic values of $N_{\rm X}$ and $T_{\rm X}$ we arrive at $N_{\rm LISA}\sim1-10$, about an order of magnitude smaller than predicted by dynamical simulations.

\subsection{NS-NS Binaries}

One NS-NS binary (M15C) has been detected directly as a binary radio pulsar, out of a total of $\sim40$ radio pulsars known in all Galactic globular clusters. A proper statistical analysis taking into account the various selection effects in pulsar surveys is well beyond the scope of this paper. However, since the dynamical processes that created M15C are thought to happen quite generically in all dense clusters (see, e.g., Phinney and Sigurdsson 1991), and since selection effects, if anything, tend to make pulsars in close binaries more difficult to detect than other types of radio pulsars (Johnston and Kulkarni 1991), it may not be unreasonable to estimate the total number of NS-NS binaries in globular clusters as $\sim N_{\rm NS}/40$, where $N_{\rm NS}$ is the total number of NS in clusters. A strict lower limit on $N_{\rm NS}$ comes from estimates of the total number of active radio pulsars in clusters, giving $N_{\rm NS} > 10^4$ (see, e.g., Kulkarni \etal\ 1990). An upper limit comes from dynamical models of globular clusters suggesting that as much as $\sim5\%$ of the total mass of a cluster may be in the form of $1.4\,M_\odot$ dark remnants (some of which may be massive WDs, which, for our purposes, are indistinguishable from NS; see, e.g., Murphy \etal\ 1998). Taking again the total mass in globular clusters $\sim10^{7.5}\,M_\odot$ (Prince \etal\ 1991), this gives $N_{\rm NS} < 10^6$. Adopting the coalescence time of M15C, $t_{\rm coal}=2\times10^8\,$yr, as typical, and using $T_{\rm LISA}= 3 \times 10^6\,$yr from eq.~1 for two $1.4\,M_\odot$ NS, we get a number $\sim 1-100$ NS-NS binaries in the LISA band, in rough agreement with the predictions of dynamical simulations.

\section{Detection Efficiency}

From the above estimates, we see that there may be many compact binaries in the Galactic globular cluster system that will be detectable GR sources for LISA. In order to determine the efficiency of LISA for detecting these sources throughout the globular cluster system, we calculated the S/N ratio for a range of binary masses and frequencies from each globular cluster in the system. We begin by modeling the strain amplitude measured by LISA. A stationary detector will measure a strain amplitude found using the point-mass quadrupole formula for a circularized binary given by
\begin{equation}
h(t) = \frac{\sqrt{3}}{2}(A_+F^+\cos{(2\pi ft)} + A_{\times}F^{\times}\sin{(2\pi ft)}),
\end{equation}
where $F_+$ and $F_{\times}$ represent the sensitivity of the detector to each polarization. The amplitudes are given by
\begin{eqnarray}
A_+ = \frac{G^{5/3}}{c^4d}M_{\rm chirp}^{5/3} f^{2/3}(1+\cos^2{i}) \\
A_{\times} = -2 \frac{G^{5/3}}{c^4d}M_{\rm chirp}^{5/3} f^{2/3} \cos{i}.
\end{eqnarray}
The angle of inclination, $i$, is the angle that the line of sight makes with the orbital plane of the binary. Since LISA is in orbit about the sun and its orientation changes, the actual signal received by LISA is given by
\begin{equation} h(t) = \frac{\sqrt{3}}{2} A(t) \cos{\left[\int^t2\pi f(t^{\prime}) dt^{\prime} + \varphi_{p}(t) + \varphi_D(t) + \varphi_0 \right]}.
\end{equation}
The Doppler shift in frequency due to LISA's motion about the sun is incorporated in $\varphi_D(t)$, and the changing orientation of LISA with respect to the source is incorporated in both $\varphi_p(t)$ and $A(t) = \sqrt{(A_+F^+)^2+(A_{\times}F^{\times})^2}$. The exact form of $\varphi_D$, $\varphi_p$, $F^+$, and $F^{\times}$ can be found in Cutler (1998). The S/N ratio $\rho$ is then found from \begin{equation} \rho^2 = 4 \sum_{\alpha}\int_0^{\infty}\tilde{h}_{\alpha}^*(f) \tilde{h}_{\alpha}(f)\frac{df}{S_n(f)}, \end{equation} with the sum over $\alpha$ representing the sum over both semi-independent arms of LISA and the Fourier transform $\tilde{h}$ defined as \begin{equation} \tilde{h}_{\alpha}(f) = \int_{-\infty}^{\infty} \rme^{2\pi \rmi f t}h_{\alpha}(t) \,dt. \end{equation}

Using globular cluster position and distance data from Harris (1996), we determined the S/N ratio for NS-WD binaries with WD masses in the range of 0.1 to 0.4 $M_{\odot}$ and frequencies ranging from $1\,$mHz up to Roche lobe overflow, using the expression of Rappaport \etal\ (1987), $f_{\rm max} = 43.3\,{\rm mHz}\, (M_{\rm WD}/M_{\odot})$. We note that the signal strength for a binary with $i = 0$ is about twice that for one with $i = \pi /2$. The globular clusters were then classified by the strength of the detected signal for binaries with favorable or unfavorable angles of inclination. There are 7 clusters with some binaries having $i = \pi / 2$ and  $\rho \geq 5$ and 17 for which only binaries with $i = 0$ have $\rho \geq 5$. The binaries associated with these clusters would have sufficient angular resolution to clearly identify the binary with its cluster. There are 72 clusters with $i = \pi/2$ and $\rho \geq 2$ and 36 with  only $i = 0$ and $\rho \geq 2$. These clusters could house binaries whose angular resolution would be a few square degrees, and the identification of the binary with its host cluster would be a little more suspect. There are only 15 clusters which are too distant to detect any binaries with $\rho \geq 2$.

\section{Conclusion}

There is ample, direct and indirect, evidence for the existence of a population of compact binaries in globular clusters that will be detectable GR sources for LISA. The observed populations of X-ray binaries and millisecond radio pulsars clearly suggest a rich and interesting variety of dynamical processes that should produce large numbers of compact binaries. A variety of estimation techniques, both theoretical and semi-empirical, are in rough agreement to predict as many as $\sim100$ NS-NS, NS-WD, and WD-WD binaries detectable by LISA in the Galactic globular cluster system. Although there are very large uncertainties in the exact numbers, the existence of these populations is not in doubt. LISA will be able to probe about 90\% of Galactic globular clusters to search for binaries with orbital periods below 2000$\,$s. Measurements of this population of binaries will provide valuable insight into the dynamical evolution of dense star clusters and the role that binaries play in them.

\ack

M.J.B. acknowledges support from NASA EPSCoR grant NCCW-0058 and Montana EPSCoR grant NCC5-240. S.P.Z. was supported by NASA through Hubble Fellowship grant HF-01112.01-98A awarded by the Space Telescope Science Institute, which is operated by the Association of Universities for Research in Astronomy. F.A.R. acknowledges support from NSF Grants AST-9618116 and PHY-0070918, and from NASA ATP Grant NAG5-8460. 

\References
\item[] Applegate J H and Shaham J 1994 {\it ApJ} {\bf 436} 312
\item[] Bailyn C D  1995 {\it ARAA} {\bf 3}, 133
\item[] Benacquista M 1999 {\it ApJ} {\bf 520} 233
\item[] Binney J and Tremaine S 1987, Galactic Dynamics (Princeton: Princeton University Press) 
\item[] Camilo F, Lorimer D R, Freire P, Lyne A G and Manchester R N 2000 {\it ApJ} {\bf 535} 975 
\item[] Cutler C 1998 {\it Phys. Rev} D {\bf 57} 7089
\item[] Davies M B 1995 {\it MNRAS} {\bf 276} 887
\item[] Deutsch E W, Margon B and Anderson S F 2000 {\it ApJL} {\bf 530} L21 \item[] Harris W E 1996 {\it AJ} {\bf 112} 1487
\item[] Hils D and Bender P L 2000 {\it ApJ} {\bf 537} 334
\item[] Hut, P, \etal\ 1992 {\it PASP} {\bf 104} 981
\item[] Johnston H M and Kulkarni S R 1991 {\it ApJ} {\bf 368} 504
%\item[] Johnston H M and Verbunt F 1996 {\it AA} {\bf 312} 80
%\item[] Joshi K J, Rasio F A and Portegies Zwart S 2000 {\it ApJ} {\bf 540} 969
\item[] Kulkarni S R, Narayan R and Romani R W 1990 {\it ApJ} {\bf 356} 174
\item[] Murphy B W, Moore C A, Trotter T E, Cohn H N and Lugger P M 1998 {\it AAS} {\bf 193} 60.01
\item[] Phinney E S and Sigurdsson S 1991 {\it Nature} {\bf 349} 220
\item[] Portegies Zwart S F and McMillan S L W 2000 {\it ApJL} {\bf 528} L17
\item[] Prince T A, Anderson S B, Kulkarni S R and Wolszczan A 1991 {\it ApJ} {\bf 374} L41
\item[] Rappaport S, Nelson L A, Ma C P and Joss P C 1987 {\it ApJ} {\bf 322} 842
\item[] Rasio F A 2000 in {\it Dynamics of Star Clusters and the Milky Way} eds. R. Spurzem \etal (ASP Conference Series) [astro-ph/0006205]
\item[] Rasio F A, Pfahl E D and Rappaport S 2000 {\it ApJL} {\bf 532} L47
\item[] Schneider R, Ferrari V, Matarrese S, and Portegies Zwart S 2000 {\it MNRAS} in press {\it Preprint} astro-ph/0002055
\item[] Tavani M 1991 {\it ApJ} {\bf 366} L27
\endrefs

\end{document}